# Controlled Intracellular Delivery of Single Particles in Single Cells by 3D Hollow Nanoelectrodes

Jian-An Huang [a], Valeria Caprettini [a,b], Yingqi Zhao [a], Giovanni Melle [a,b], Nicolò Maccaferri [a], Matteo Ardini [a], Francesco Tantussi [a], Michele Dipalo [a], Francesco De Angelis [a,*]

[a] Istituto Italiano di Tecnologia, Via Morego 30, 16163 Genova, Italy
[b] DIBRIS, University of Genoa, Via all'Opera Pia 13, 16145 Genova, Italy
* francesco.deangelis@iit.it.

*Supporting Information included*

**Abstract:** We present an electrophoretic platform based on 3D hollow nanoelectrodes capable of controlling and quantifying the intracellular delivery of single nanoparticles in single selected cells by surface-enhanced Raman spectroscopy (SERS). The gold-coated hollow nanoelectrode has a sub-femtoliter inner volume that allows the confinement and enhancement of electromagnetic fields upon laser illumination to distinguish the SERS signals of a single nanoparticle flowing through the nanoelectrode. The tight wrapping of cell membranes around the nanoelectrodes enables effective membrane electroporation such that single gold nanorods are delivered into a living cell with a delivery rate subject to the applied bias voltage. The capability of the 3D hollow nanoelectrodes to porate cells and reveal single emitters from the background under live flow is promising for the analysis of both intracellular delivery and sampling.

**Significance:** The delivery of molecules into the intracellular compartment is one of the fundamental requirements of the current molecular biology. However, the possibility of delivering a precise number of nano-objects (nanoparticles, proteins, genic materials) with single-particle resolution is still an open challenge. Here, we show that single nano-objects can be delivered into cells cultured in vitro by an electrophoretic approach based on 3D hollow nanoelectrodes combined with surface-enhanced Raman scattering, which enables real-time counting of individual nanoparticles. By a simple refinement of the platform, it is possible to realize one nanoelectrode per cell, which would be ideal for use in the emerging field of single-cell technology and the on-chip analysis of the content extracted from cells, including proteins, DNA and miRNA.

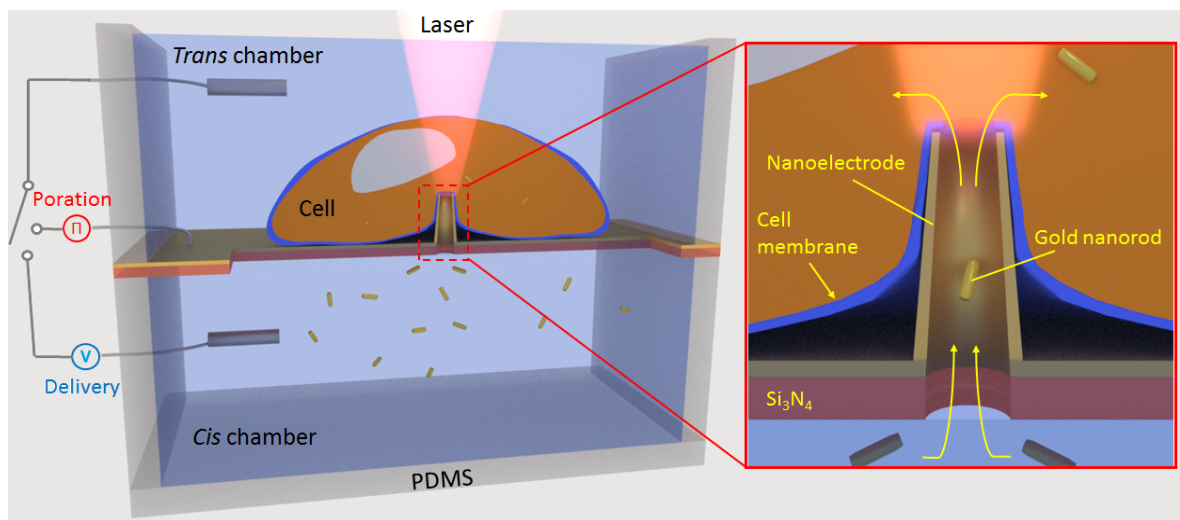

Figure 1. Schematic representation of the 3D hollow nanoelectrode device for single-particle intracellular delivery. The cell is tightly wrapped around the gold-coated hollow nanoelectrode and is first electroporated by a pulsed voltage. Then, the nanorods originally in the cis chamber are delivered into the cell through the hollow nanoelectrode by a DC potential between the two Pt wire electrodes. Inset: a laser excites the Raman signals of the delivered nanorods for counting the number of delivered nanorods.

## Introduction

The intracellular delivery of nanoparticles, such as quantum dots and gold nanoparticles, is widely used in proteomics, drug delivery, and single-cell studies.[1-10] Nanoparticle endocytosis usually leads to nanoparticle aggregation in endosomal vesicles and nanoparticle attachment to the cell membrane.[11] However, these vesicles can prevent the trapped nanoparticles from approaching targeted organelles or molecules.[12] Additionally, nanoparticle aggregates much larger than single nanoparticles could distort molecular behavior in mechanistic studies, such as those of intracellular transport by motor proteins.[13-20] Therefore, physical delivery methods for injecting single particles into the cytoplasm of living cells are highly desired.

Although many different approaches have been developed for cells cultured in vitro,[21] some important limitations, such as poor dosage control, remain.[22] Among them, the development of quantitative methods has remained difficult. In particular, the possibility of delivering a precise number of nano-objects with single-particle resolution is an ongoing issue. Moreover, the ability to target single selected cells within a large population would be of additional value in the emerging field of single-cell biology, which aims to discover characteristics of individual cells that are hidden in experiments performed using large cell numbers.

In recent years, different physical methods based on nanopores[23,24] and hollow nanotube systems (or nanostraws)[25-27] have been developed as reliable means of delivery with high cell viability. However, these methods have not overcome the aforementioned limitations.

In this work, we show that single nano-objects can be delivered into cells cultured in vitro by an electrophoretic approach combined with surface-enhanced Raman scattering, which enables the counting of individual nanoparticles in real time. The method is based on plasmonic hollow nanotubes that simultaneously act as nanoelectrodes for cell electroporation and particle delivery (nanochannels) and as plasmonic antennas for Raman signal enhancement. The concept is represented in Figure 1. Hollow nanotubes are fabricated on a $Si_3N_4$ substrate embedded in a polydimethylsiloxane (PDMS) chamber to separate a trans chamber from a cis chamber. The gold-coated hollow nanotube interfacing with the cell membrane acts as a nanoelectrode to generate electropores by a pulsed voltage. Then, DC potential is applied to two Pt wire electrodes in both chambers to deliver nanorods from the cis chamber to the electroporated cell through the hollow nanoelectrodes. The optical energy can be confined inside the hollow nanotube[28] such that upon laser illumination, single nanorods entering the nanoelectrodes can be well distinguished from the background in the cis chamber.

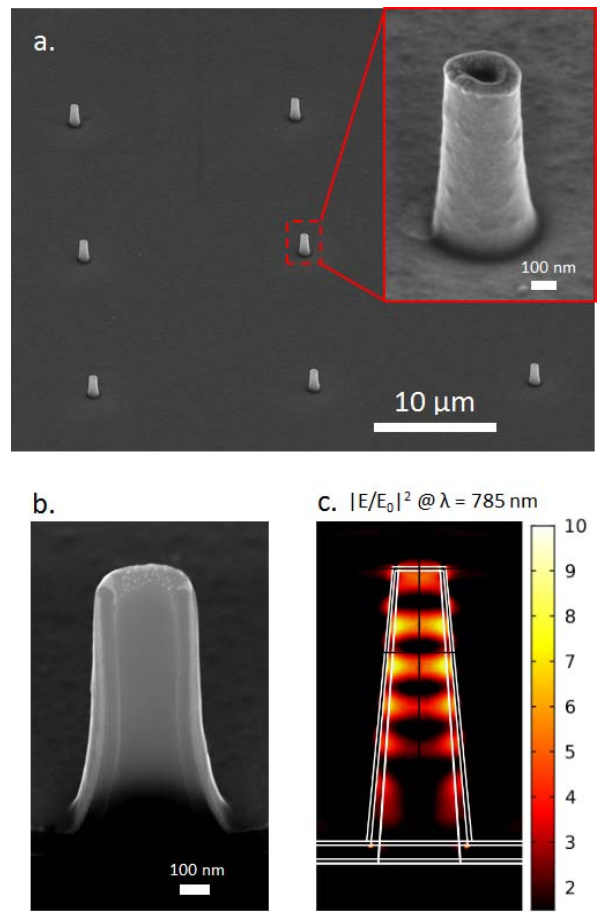

Figure 2. SEM images of 3D hollow nanoelectrode array on $Si_3N_4$. Inset: magnified SEM image of a single nanoelectrode (a). The nanoelectrode has a hollow volume and is 300 nm in diameter and 2 ftm in length (b). Simulated electromagnetic field intensity distribution of an illuminated nanoelectrode (enhancement factor between 5 and 10) (c).

## Results and Discussion

We first assessed the performance of the 3D hollow nanoelectrodes by quantifying single-particle translocation through the hollow nanoelectrodes in phos-phate-buffered saline without cells. Raman-tagged gold nanorods 25 × 90 nm in size (Supporting Information Figure S1) were used for electrophoretic translocation. Under 785-nm illumination, single nanorods exhibited stable Raman spectra (Supporting Information Figure S2) in which the signal-to-baseline intensity of the Raman band at 593 cm$^{-1}$ was used as the signal for counting the translocated nanorods.

The hollow nanoelectrodes covered by a 30-nm-thick gold layer had an inner diameter of 300 nm and a length of 2 μm, as shown in Figure 2a and 2b. When illuminated with a 785-nm laser, the electromagnetic field intensity was enhanced by a factor of up to 10-fold the intensity of the incident field, as shown in Figure 2c, and as previously demonstrated.[28]

The inner volume of the nanotube was approximately 0.14 fL. However, the experimental detection volume is expected to be even smaller because, as shown in Figure 2c, the plasmonic field is accumulated in a total volume that is smaller than that of the nanotube. By collecting the Raman signal using an objective with a high numerical aperture (NA = 1) focused at the nanotube tip, we detected only the nanorods translocating into the nanotube, whereas the nanorods dispersed in solution in the cis chamber contributed to a very low background noise level.

To allow a single nanorod to translocate through the nanotube (i.e., to prevent the coincidence of two particles in the same time window), the concentration of nanorods dispersed in solution must be carefully adjusted.[29] We used Poisson statistics to calculate the optimal concentration (see Methods for details).[30] According to the calculations, a concentration of 10$^{11}$ particles per mL leads to 0.014 nanorods diffusing inside the hollow nanoelectrode, on average. The probability of 0, 1 and 2 nanorods diffusing in the hollow nanoelectrode was 0.986, 1.38 × 10$^{-2}$ and 9.66 × 10$^{-5}$, respectively. Since the probability of having 2 nanorods passing through the nanochannel simultaneously was very low (<10$^{-4}$), we assumed that the recorded Raman signal was always due to a single nanorod.

### Electrophoretic Translocation

DC voltages ranging from -0.5 to -2 V were used for the electrophoretic translocation since the nanorods were negatively charged. The DC potential was applied between two Pt wire electrodes that were separated by a distance of approximately 15 mm in the cis and trans chambers. Single-particle translocations were demonstrated as bursts in a time trace of the intensity changes of the nanorod Raman band at

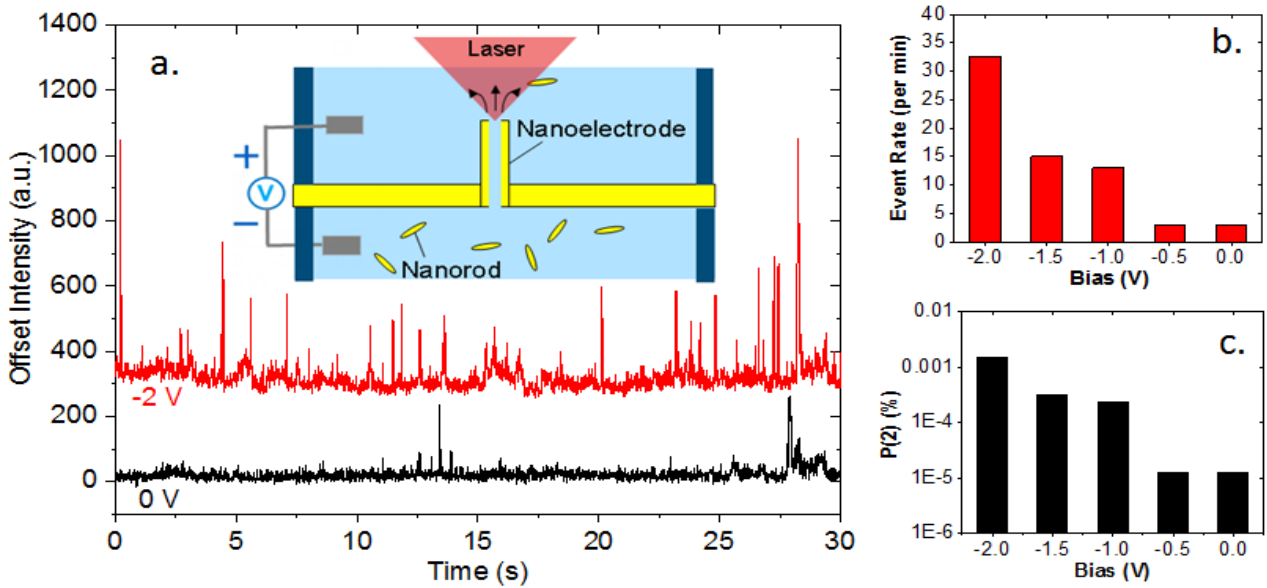

Figure 3. Electrophoretic translocation of nanorods through the hollow nanoelectrodes without cells. (a) Time traces of electrophoretic translocation at a DC bias of -2 V (red curve) and 0 V (black curve, diffusion regime). Inset: schematic of the electrophoretic translocation without cells. (b) Measured event rates of nanorod translocation; a burst with a signal-to-noise intensity ratio no less than 3 was considered an event. (c) Probability of the coincidence of 2 nanorods in flow during electrophoretic translocation.

593 cm$^{-1}$, as shown in Figure 3a.

A burst with a signal-to-noise intensity ratio no less than 3 was considered a single-particle translocation event. In the case of spontaneous diffusion (no bias applied), only 2 events occurred within 30 s. In contrast, the number of events increased significantly at a bias of -2 V. Thus, the event rate increased with the applied potential, indicating that more single nanorods were translocated in a certain time window (Figure 3b). The average translocation time obtained was 57 ms at a bias of -2 V (Supporting Information Figure S3). Although the translocation time depended on many parameters, such as the ratio of the nanorod size to the nanoelectrode diameter, the translocation time obtained is on the same order as those from previous reports of single-particle translocation through nanopores,[31-34] confirming that the observed events were related to single-particle translocation.

A critical parameter for ensuring single-particle translocation is the low probability of the coincidence of more than one particle in flow. Unlike the diffusion case, the probability calculated by the Poisson statistics for particles in flow[35] considers the irreversible flow of the nanoparticles with a flow rate (or event rate) and an exposure time as the detection time window (see Methods for details). At an exposure time of 10 ms and a measured event rate of 32 per minute, the probability of translocating 2 nanorods simultaneously at a bias of -2 V was calculated as $1.46 \times 10^{-5}$. The probabilities at other bias voltages were even smaller, as shown in Figure 3c. Thus, the electrophoretic translocation of a single nanorod was confirmed as the most likely cause for event detection. Compared with particle diffusion, electrophoresis prevents translocated nanorods from returning to the nanoelectrodes. Moreover, this approach provides more cognizant control of the translocation rate. Here, the electrophoretic voltage of our hollow nanoelectrode system was optimized with the nanorod concentration for the efficient intracellular delivery of single nanorods, as shown below.

### Intracellular Delivery

To demonstrate intracellular delivery, NIH-3T3 cells were cultured in the trans chamber to allow cell growth on the hollow nanoelectrodes with tight membrane wrapping (Figure 4). Together with two Pt wire electrodes for translocating the nanorods, a ca-ble was connected to the gold layer of the hollow nanoelectrodes for cell membrane electroporation. The membrane was porated by applying a peak-to-peak pulsed voltage of 3 V for 10 s with pulse length of 100 fts and a frequency of 20 Hz between the Pt wire electrode in the trans chamber and the hollow nanoelectrodes. After the electropores were generated in the cell membrane, electrophoretic delivery of the nanorods was conducted with DC voltage (-0.5 to -2 V) between the two Pt wire electrodes in the trans and cis chambers. Gold nanorods with 10 x 40 nm in size were used to facilitate delivery through the small electropores. A time trace of the electroporation and delivery exhibited delivery events at a bias of -2 V with an event rate of 4 min$^{-1}$ after membrane electroporation, as shown in Figure 5a.

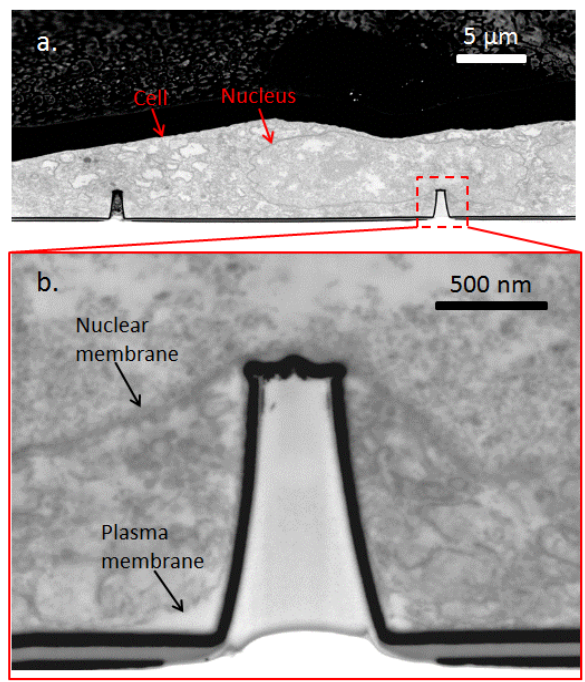

Figure 4. Cross-sectional SEM image of a cell cultured on the nanoelectrodes (a). A magnified SEM image showing that the cell membrane is tightly wrapped around the nanoelectrode (b).

That fact that no bursts were observed from a bias of -0.5 to -1.5 V suggests that the electroporated cell membrane presented many barriers to the electrophoretic delivery. One such barrier could be that the resistance of the cell membrane decreased the electrophoretic voltage, even if the cell membrane was electroporated.[36] In addition, the transient electro-pores continuously shrink after electroporation,[37, 38] leading to a low event rate. Such an event rate corresponds to the probability of coincidence in flow as small as $10^{-7}$, ensuring single-particle intracellular delivery.

To assess the nanorod delivery, we examined the cells by Raman mapping and analyzing the Raman band at 593 cm$^{-1}$. The distribution and transport of the nanorods are shown in merged images in Figure 5b and 5c, in which the colored dots indicate the signal-to-baseline Raman intensity of the nanorods. In previous reports on the endocytic uptake of nanoparticles, the Raman signals of intracellular nanoparticles were colocalized with black dots in bright-field optical images that corresponded to intravesicular nanoparticle aggregates.[39, 40] Our case is in strong contrast with these reports because no black dots were observed in the bright-field images overlapping the colored dots. Since it takes at least 2 hours for motor proteins to capture and aggregate single intracellular nanoparticles,[41, 42] the colored dots should correspond to single nanorods, which could not be resolved by the bright-field 60× objective.

The number of colored dots in the Raman maps that represent delivered nanorods can hardly be equal to the number of bursts of the time trace. On the one hand, the Raman mapping was performed spot-by-spot by mechanically moving the sample stage. As each spot required 300 ms to map, the whole mapping process was usually completed in 3 to 5 minutes, which was too slow to trace the real-time distribution of the nanorods. On the other hand, the delivered nanorods could have moved from the focal plane to the upper interior of the cell during the Raman mapping. Nevertheless, the temporal limitation due to the stage-scanning Raman microscope can be readily overcome using a laser-scanning Raman microscope.[43, 44]

The continuous colocalization of the colored dots with some nanoelectrode positions in the Raman maps suggests that the nanorods were accumulated inside the nanoelectrodes. This accumulation could have occurred because the electropores were too small to allow nanorod passage. When the accumulated nanorods were clogged inside the nanoelectrode, they could be moved back into the cis chamber by applying a positive DC bias (Supporting In-formation Figure S4). This control of reversible nanorod movement through the nanoelectrodes could be a way for the hollow nanoelectrodes to extract objects from porated cells and perhaps be applied for intracellular sampling.[45]

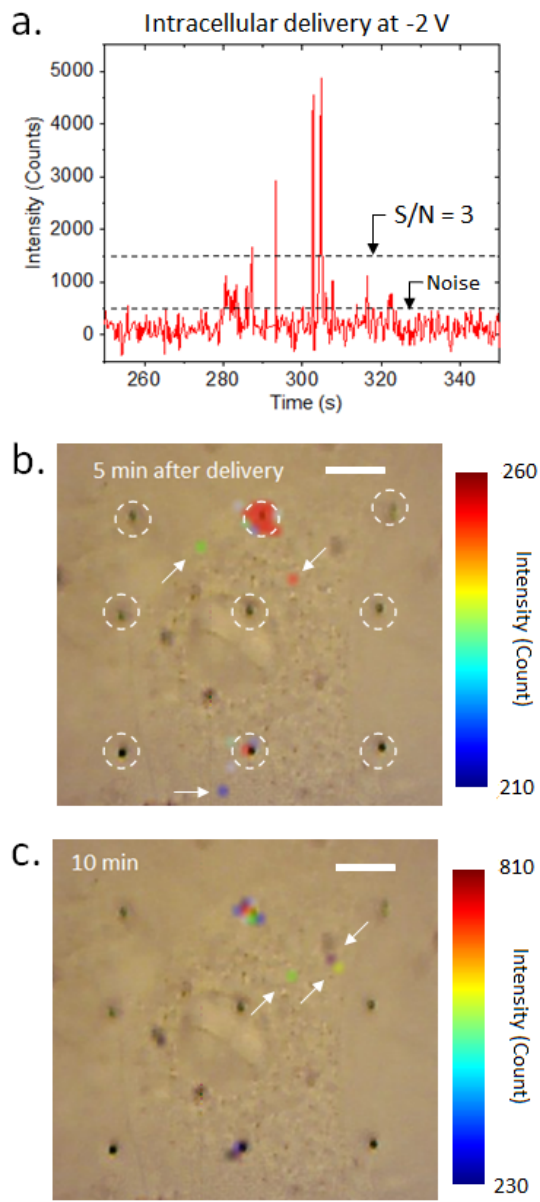

Figure 5. A time trace of the electrophoretic intracellular delivery of nanorods at a bias of -2 V after electroporation (a). Optical images merged with Raman maps of the nanorods. White dotted circles are the locations of the nanoelectrodes. White arrows indicate the nanorods delivered into one cell. The colored dots represent the Raman intensity of the nanorods, as indicated by the colored bars. Intracellular delivery of the nanorods after 5 min (b) and 10 min (c). All scale bars are 10 μm.

## Conclusion

We demonstrated the electrophoretic intracellular delivery of nanorods with the capability of controlling and detecting single events, thus providing a method for accurate quantitative delivery. The platform is based on multifunctional plasmonic hollow nanotubes that can work as i) nanoelectrodes for cell electroporation, ii) nanochannels for electrophoretic delivery, and iii) plasmonic antennas to enhance the optical signals of nano-objects translocating through the channel. The tight wrapping of the cell membrane around the nanoelectrodes allowed the generation of electropores large enough to allow nanorod passage while still preserving the membrane adhesion after electroporation and preventing nanorod leakage. By a simple refinement of the platform, it is possible to realize one nanoelectrode per cell for the discrimination of cells that have and have not received a nano-object, which would be ideal for use in the emerging field of single-cell technology. Finally, as hollow nanotubes have been demonstrated to extract cytosolic context from living cells,[45] in the future, such a platform could be used for the real-time monitoring and on-chip analysis of the content extracted from cells, including proteins, DNA and miRNA.[46]

## Methods

**Materials.** Raman-tagged gold nanorods dispersed in deionized water were purchased from Nanopartz Inc. (Loveland, CO, USA) with Nile blue A (NBA) as the Raman reporter and stabilized by carboxyl groups. The gold nanorods were either 10 × 40 nm or 25 × 90 nm in size, with transverse plasmonic resonance at a wavelength of 510 nm and longitudinal plasmonic resonance at a wavelength of 780 nm. The zeta potential at pH = 7 and the concentration of the 10 × 40-nm gold nanorods was -18 mV and 4 × $10^{13}$ particles per mL, respectively, and that of the 25 × 90-nm nanorods was -15 mV and 4 × $10^{12}$ particles per mL, respectively.

**Device fabrication.** To fabricate the 3D hollow nanoelectrodes, S1813 photoresist (Shipley) was spin-coated on a 1 x 1-cm $Si_3N_4$ membrane at 4000 rpm for 1 min and soft baked at 95°C for 5 min. After sputtering a 7-nm-thick titanium and a 20-nm-thick gold layer on the back of the $Si_3N_4$ membrane, focused ion beam milling (FIB, FEI Helios NanoLab 650 DualBbeam) at a voltage of 30 keV and a current from 0.23 to 2.5 nA was used to drill hole arrays in the back of the Ti/Au-coated $Si_3N_4$ sample. Different FIB currents correspond to different nanotube inner diameters. Then, the sample was ashed by oxygen plasma at 100 W for 2 min to smooth the photoresist and was then developed in acetone for 2 min to form polymer nanotube arrays. Then, the nanotube arrays were thinned down by oxygen plasma at 100 W for 2 min. An alumina layer of 5 nm was deposited on the back of the sample by atomic layer deposition (Oxford Instruments) to neutralize the surface charge. After being coated with a 7-nm-thick Ti layer and a 30-nm-thick gold layer by sputtering at a 45° tilt angle with rotation to ensure uniform coating, the sample was annealed on a hot plate at 200°C in the air for 1 hour and allowed to cool naturally. The as-made nanoelectrodes were attached with a cable by silver paste and embedded in a microfluidic chamber made from polydimethylsiloxane (PDMS, Dow Corning SYLGARD 184 silicone elastomer) at 60°C for approximately 40 min.

**Cell culture.** We used NIH-3T3 cells for the delivery experiments. Before seeding the cells, the devices were irradiated with UV rays for 30 min in a laminar-flow hood to sterilize them. The devices were treated O.N. with complete DMEM to saturate the PDMS chamber. Then, NIH-3T3 cells were seeded on the devices at a concentration of 0.8 × $10^4$ cells/cm$^2$ and incubated at 37°C in a 5% $CO_2$ atmosphere for 24 h in DMEM with 1% pen/strep antibiotic and 10% fetal bovine serum (Sigma Aldrich) before the experiments were performed.

**Raman measurements.** Raman measurements were obtained by a Renishaw inVia Raman spectrometer with a Nikon 60 × water immersion objective with a 1.0 NA delivering a 785-nm laser with a power of approximately 3.27 mW. Intracellular nanoparticle delivery was measured using an Andor EMCCD camera (DU970P-BVF) integrated into the Renishaw spectrometer with an exposure time of 10 ms. Cell mapping was conducted with the Renishaw CCD camera at an exposure time of 300 ms and a step of 1 ﬂm.

**Poisson statistics for particle diffusion.** When an average of <N> nanoparticles are diffusing in a given volume, the probability of having m nanoparti-

cles at any time in the volume can be calculated by the Poisson statistics for diffusion:[30]

$$P_m = \frac{\langle N \rangle^m}{m!} \exp(-\langle N \rangle) \qquad (1)$$

**Poisson statistics for particles in flow.** When nanoparticles are in flow in a given volume at a rate of c, the probability of having n nanoparticles in the volume at time {t can be calculated by the Poisson statistics in flow:[35]

$$P(n) = \frac{(c_\Delta t)^n}{n!} \exp(-c_\Delta t) \qquad (2)$$

## Acknowledgments

We thank Dr. Xavier Zambrana Puyalto and Dr. Rosario Capozza for their valuable discussions. The research leading to these results was funded by the European Research Council under the European Union's Seventh Framework Programme (FP/2007-2013)/ERC Grant Agreement no. [616213] and CoG: Neuro-Plasmonics.

## Author Contributions

F.D.A. conceived and supervised the work. J.A.H. and M.A. prepared the nanoparticles. J.A.H., V.C. and Y.Z. designed and fabricated the devices. V.C. and G.M. cultured the cells. J.A.H., V.C., F.T. and M.D. designed the optical-electrical setup. J.A.H. and V.C. performed the Raman measurements. J.A.H. and Y.Z. analyzed the data. N.M. performed the electromagnetic simulations. All authors contributed to the manuscript preparation.

# Supporting Information

# Controlled Intracellular Delivery of Single Particles in Single Cells by 3D Hollow Nanoelectrodes

*Jian-An Huang*[a]*, Valeria Caprettini*[a,b]*, Yingqi Zhao*[a]*, Giovanni Melle*[a,b]*, Nicolò Maccaferri*[a]*, Matteo Ardini*[a]*, Francesco Tantussi*[a]*, Michele Dipalo*[a]*, Francesco De Angelis*[a]*, \**

[a] Istituto Italiano di Tecnologia, Via Morego 30, 16163 Genova, Italy
[b] DIBRIS, University of Genoa, Via all'Opera Pia 13, 16145 Genova, Italy

\* francesco.deangelis@iit.it

Supplementary Figures

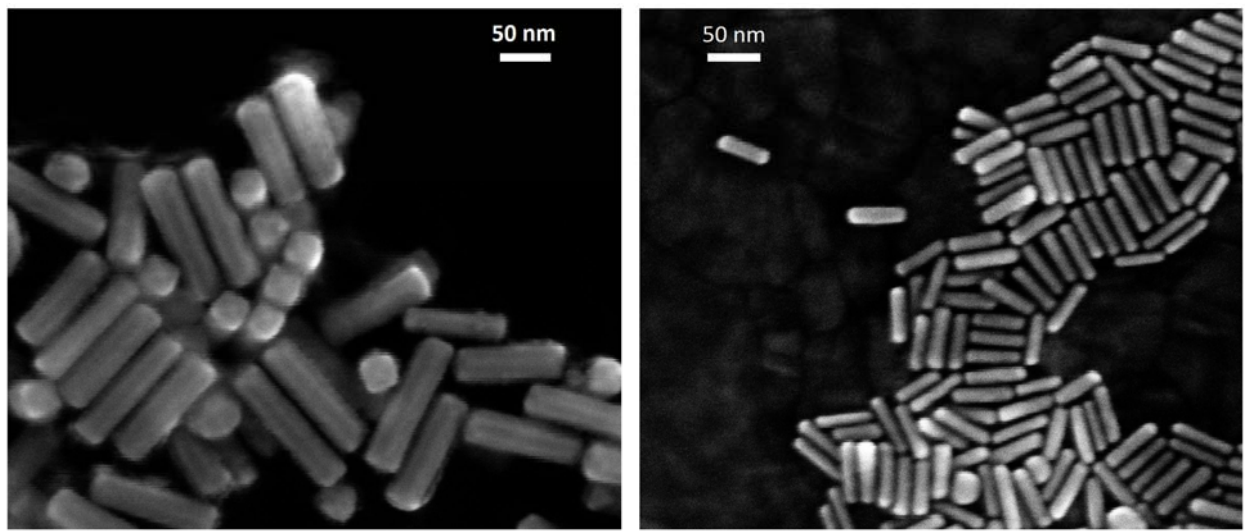

Figure S1. SEM images of the Raman-tagged gold nanorods with sizes of 25 × 90 nm (left) and 10 × 40 nm (right).

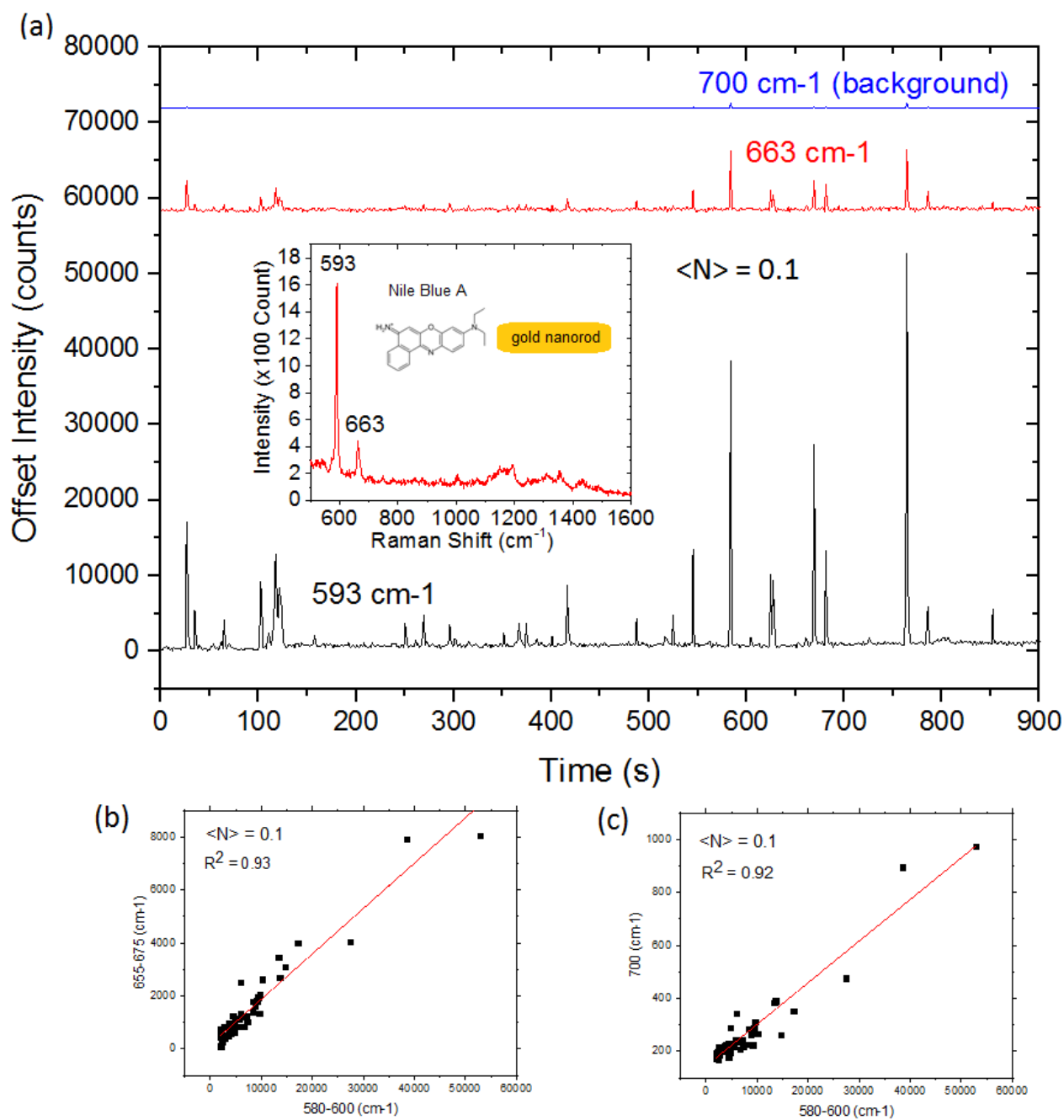

Figure S2. Raman spectrum of the Raman tag, Nile Blue A, adsorbed on the gold nanorods in which the 593 cm-1 and 633 cm-1 bands are selected for evaluation of the nanorod aggregation (Inset). Time traces of signal-to-baseline intensity of different Raman peaks and baseline (700 cm-1) of 0.1 nanorod on average diffusing in a detection volume (Φ 1.2 × 5 μm) of the 60 × water immersion objective with N.A. = 1.0. (a). The corresponding correlation of the Raman peaks between 593 cm-1 and 663 cm-1 (b) or between 593 cm-1 and the baseline at 700 cm-1 (c) in which the $R^2$ is the coefficient of correlation and <N> is the average number of nanoparticle in the detection volume. The high correlations between the Raman peaks or baseline indicate that no nanorod aggregation exists and the spectra of single nano-rod in flow are stable. [1,2]

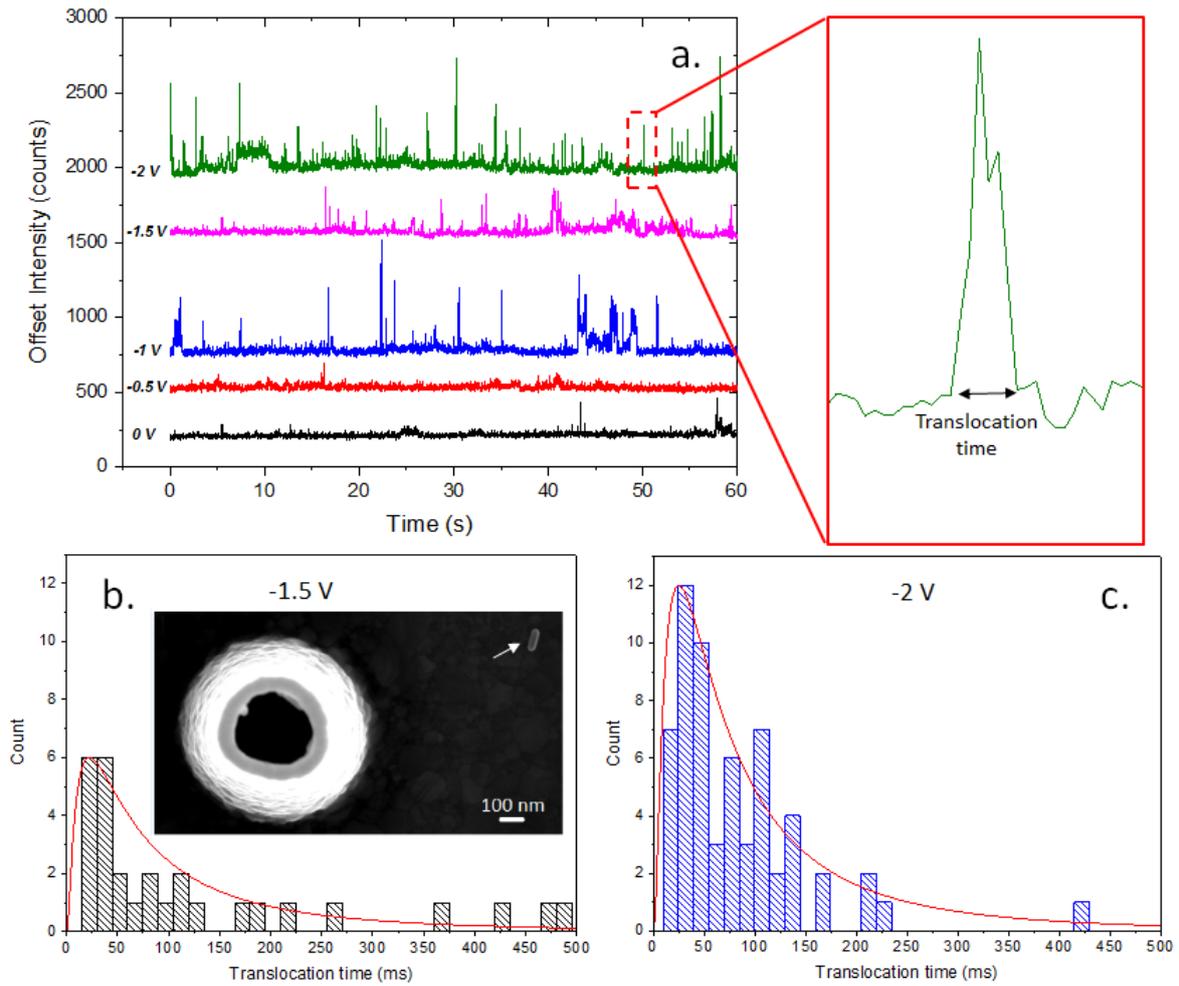

Figure S3. Time traces of the nanorods (25 × 90 nm) translocating through the nanoelectrodes without cells under different electrophoresis bias voltage; Inset is the definition of the translocation time of a burst (a). Histograms of the corresponding translocation times fitted by lognormal probability functions (red curves) to extract average translocation times: 77 ms at -1.5 V bias (b) and 57 ms at -2 V bias (c), respectively. The SEM image in the Inset of (b) indicates the translocated nanorods. The bursts in the histograms are selected with thresholds: intensity signal-to-noise ratio no less than 3 and the translocation times limited from 10 to 500 ms.

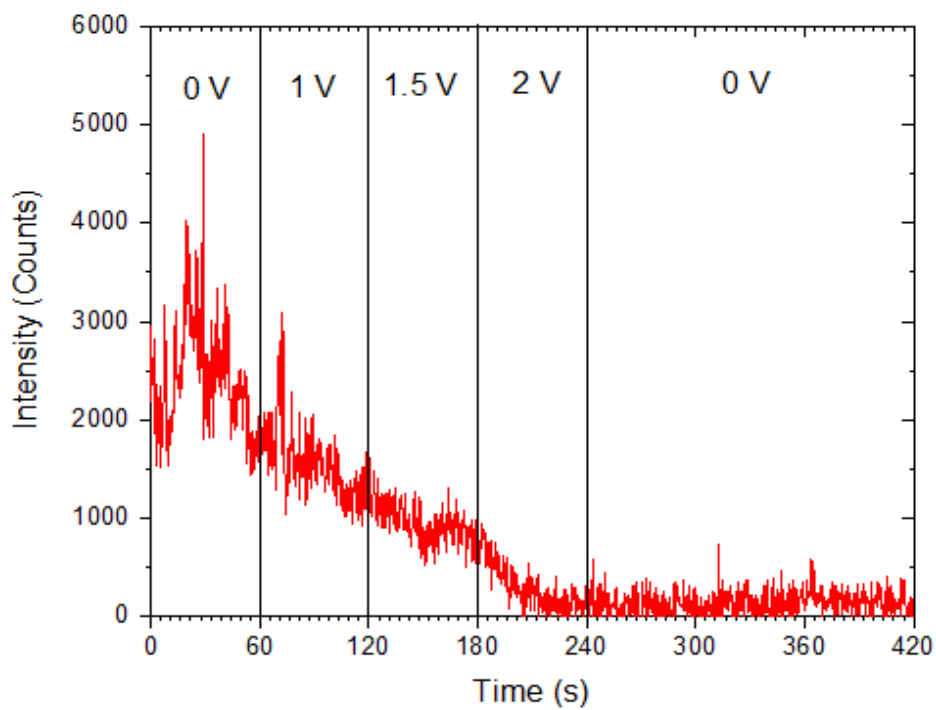

Figure S4. The time trace of nanorods (25 × 90 nm) clogging in a nanoelectrode. The nanorods were repelled by positive bias electrophoresis at different voltages. The fact that intensity decreased to near zero suggested that the nanorods were cleared out from the nanoelectrodes.